\documentclass[conference, nonieee]{IEEEtran}
\IEEEoverridecommandlockouts
\usepackage{cite}
\usepackage{amsmath,amssymb,amsfonts}
\usepackage{algorithmic}
\usepackage{graphicx}
\usepackage{textcomp}
\usepackage{xcolor, soul, enumitem}
\def\BibTeX{{\rm B\kern-.05em{\sc i\kern-.025em b}\kern-.08em
    T\kern-.1667em\lower.7ex\hbox{E}\kern-.125emX}}
\begin{document}

\title{Exploration of Novel Neuromorphic Methodologies for Materials Applications}

% \author{\IEEEauthorblockN{Derek Gobin}
% \IEEEauthorblockA{\textit{Electrical and Computer Engineering} \\
% \textit{George Mason University}\\
% Farifax, VA, USA \\
% dgobin@gmu.edu}
% }

\makeatletter
\newcommand{\linebreakand}{%
  \end{@IEEEauthorhalign}
  \hfill\mbox{}\par
  \mbox{}\hfill\begin{@IEEEauthorhalign}
}
\makeatother

\author{\IEEEauthorblockN{Derek Gobin}
\IEEEauthorblockA{
\textit{George Mason University}\\
Fairfax, VA, USA \\
dgobin@gmu.edu}

\and
\IEEEauthorblockN{Shay Snyder}
\IEEEauthorblockA{
\textit{George Mason University}\\
Fairfax, VA, USA \\
ssnyde9@gmu.edu}

\and
\IEEEauthorblockN{Guojing Cong}
\IEEEauthorblockA{
\textit{Oak Ridge National Laboratory}\\
Oak Ridge, TN, USA \\
congg@ornl.gov}

\and
\IEEEauthorblockN{Shruti R. Kulkarni}
\IEEEauthorblockA{
\textit{Oak Ridge National Laboratory}\\
Oak Ridge, TN, USA \\
kulkarnisr@ornl.gov}
% \textit{Computer Science and Mathematics Division} \\
\linebreakand
% \and
\IEEEauthorblockN{Catherine Schuman}
\IEEEauthorblockA{
\textit{University of Tennessee - Knoxville}\\
Knoxville, TN, USA \\
cschuman@utk.edu}

\and
\IEEEauthorblockN{ Maryam Parsa}
\IEEEauthorblockA{ \textit{George Mason University} \\
Fairfax, VA, USA\\
mparsa@gmu.edu}
}

\maketitle

\begin{abstract}
Many of today's most interesting questions involve understanding and interpreting complex relationships within graph-based structures. For instance, in materials science, predicting material properties often relies on analyzing the intricate network of atomic interactions. Graph neural networks (GNNs) have emerged as a popular approach for these tasks; however, they suffer from limitations such as inefficient hardware utilization and over-smoothing. Recent advancements in neuromorphic computing offer promising solutions to these challenges. %approaches explore the application of neuromorphic computing to overcome the aforementioned issues plaguing traditional GCNs.
In this work, we evaluate two such neuromorphic strategies known as reservoir computing and hyperdimensional computing. We compare the performance of both approaches for bandgap classification and regression using a subset of the Materials Project dataset. Our results indicate recent advances in hyperdimensional computing can be applied effectively to better represent molecular graphs.
\end{abstract}

\section{Introduction}
\noindent
Graph Neural Networks (GNNs) are a popular strategy wherever data can be expressed as a graph. The materials science domain is no exception, where atomic structures naturally lend themselves to graphical representations. 
%The capability of these models for materials applications is demonstrated across various tasks within The Materials Project dataset \cite{b1}, where GCNs make up the majority of the state of the art \cite{b2}. 
This success is due to GNNs' ability to utilize the information and structure of graphs to generate feature embeddings. Their state-of-the-art performance is demonstrated across various tasks within the Materials Project dataset~\cite{b1,b2}. 

Despite these impressive results and the research attention they have brought, GNNs still suffer from a number of challenges. Like all traditional deep learning strategies, they rely heavily on the quantity and quality of data. This is particularly significant in the materials domain, where obtaining high-quality data for less common material structures can be expensive and time-consuming~\cite{b3}.
%specific information about certain structures can be limited \cite{b3}. 
Another issue is that graphs tend to be sparse with some nodes more heavily connected than others~\cite{cong2023hyperparameter}. These attributes make them inefficient on today's standard computational hardware that relies primarily on parallel processing. Finally, one of the most significant challenges faced by today's GNN researchers is over-smoothing, which can limit the expressiveness and representative power of GNN architectures. These challenges have led to GNNs being surpassed in some tasks by more standard feed-forward strategies that require handcrafted feature selection and careful design \cite{b3}. 

In recent years, neuromorphic computing has emerged as a popular area of research due to its efficiency and potential to mimic biological learning processes. Neuromorphic hardware, inherently asynchronous and optimized for handling sparse data, 
%designed to be asynchronous and handle sparsity, 
is particularly well-suited for the graph domain. Furthermore, promising learning strategies have developed for generating feature representations. Reservoirs, for example, have shown to be effective feature extractors across various modalities through the use of a large untrained layer of neurons \cite{b11}, \cite{b14}. Alternatively, hyperdimensional computing builds symbolic representations of data through projection to high dimensional spaces. These methodologies are less data intensive and computationally expensive than their deep learning counterparts, requiring much less resources to train and capable of running natively on neuromorphic hardware. To examine these recent advances for application to the materials domain, we apply reservoir and hyperdimensional computing strategies to a small subset of the Materials Project dataset \cite{b1} for bandgap classification and regression tasks.

Our key contributions are as follows: \begin{itemize}
    \item We perform an initial exploration and comparison of recent reservoir and hyperdimensional computing strategies for the representation of molecular graph structures.
    \item We introduce a novel strategy, called SSP-GrapHD, for representing molecular structures and provide preliminary results.
    \item Our results show that SSP-GrapHD reduces the mean absolute error when compared to the state-of-the-art ALIGNN approach~\cite{b5} on our data subset.
\end{itemize}

% \textcolor{red}{Personally, I would want to add some more detail about the different methods that are examined in this paper. Not too much information just enough to give the reader a rough idea. They should be able to go to the background to learn more. - Shay}

\section{Related Work}

%\textcolor{red}{This is a personal preference thing but some people like to include a short outline of the subsections within each section. Whereas some other do not. I would think about which strategy you like better}

%\subsection{GCNs}

\noindent
Graph Neural Networks (GNNs) operate through a message passing paradigm, where nodes receive information from their connected neighbors. This information encodes the features of the neighboring node and the edge connecting them.
%These messages are meant to contain information about the neighbor node and its corresponding edge.
Through this process, GNNs construct node embeddings that capture both the structure and the data of the graph. 
%This leads to the construction of various node embeddings, which together form a representation of both the structure and the data of the graph. 
While theoretically, these layers could continue to generate embeddings that capture the entire network of connections within the graph, in practice, GNNs are limited to one or two layers before over-smoothing becomes a problem. 
%The first layer generates node embeddings using nodes directly connected to the querying node, while the second layer generates node embeddings based on nodes two connections away. In theory, these layers could continue to generate node embeddings that capture the full graph structure; however, in practice, GCNs are generally limited to one or two layers before over-smoothing begins taking effect. 
Over-smoothing refers to the process where connected nodes receive similar information across the graph, leading to homogeneous representations and a loss of information \cite{b4}. For materials discovery, GNN techniques often rely on constructing line graphs to augment the atomic structure graph and utilize careful feature selection in addition to graph design \cite{b5}, \cite{b6}. %Attention mechanisms have also been incorporated to some success \cite{b7}. 

% \subsection{Neuromorphic}

Neuromorphic applications to complex graph problems have mainly focused on the text domain and citation networks \cite{b8}, \cite{b9}, \cite{b10}. %In particular, knowledge graph representation has been shown to be of interest. 
These strategies exploit specific characteristics of these domains, which are not directly applicable to the material science domain. However, the inherent efficiency and low-power nature of neuromorphic computing holds promise for overcoming limitations in processing large material science graphs.
%These strategies tend to exploit the structure of their target knowledge graph (e.g. no repeated node values) and use delays to determine the feasibility of queries \cite{b9}, \cite{b10}.

Reservoir computing, a neuromorphic strategy, utilizes a large recurrent network of untrained neurons (called a reservoir) to generate high-dimensional feature representations. These representations are then fed into a separate (usually linear) read-out layer trained for the specific classification or regression task. Reservoirs can be built with standard neurons (called echo state networks) or biologically inspired spiking neurons (called liquid state machines). In \cite{b11}, the authors explore using an echo state network for graph classification on a small subset of the MUTAG molecular dataset.  This work primarily focuses on examining the hardware implementation of a reservoir, achieving comparable results to their GNN baselines, but with limited comparison to the state-of-the-art methods.
% This effort is more of an examination of the hardware implementation of a reservoir, and little effort to compare to the state of the art is made, although the authors did note that they achieved comparable results to their GCN baselines. 

Hyperdimensional computing is a relatively new approach to computation inspired by how the brain represents information across a large number of synapses at any given moment. Data is represented through projection to a high dimensional space, where it can be manipulated via algebraic operations (e.g. binding, bundling, permutation). The specific operations used depend on the chosen type of hypervector, which can range from binary vectors to more complex tensor representations \cite{b12}. In this work, we evaluate the performance of reservoir and hyperdimensional learning algorithms within neuromorphic GNNs applied to materials science problems. We compare their performance for bandgap classification and regression tasks using the Materials Project dataset.

\section{Methodology}

\noindent
To explore our selected strategies, two tasks were identified for bandgap prediction: a simple binary classification problem (zero vs. non-zero bandgap) and a more challenging regression problem (estimating the actual bandgap value). The data used for this evaluation consisted of a selection of 54 data points from the Materials Project dataset ~\cite{b1,b2}. This selection allows for an efficient evaluation of the performance of reservoir and hyperdimensional computing for these tasks while enabling further exploration with larger datasets in future work. 

\subsection{Reservoir Computing}

\noindent
For the reservoir computing architecture, we explored a liquid state machine (LSM) due to its use of biologically plausible spiking neurons.These spiking neurons are well-suited for processing the sparse graph representations and offer potential for further efficiency gains by leveraging specialized neuromorphic hardware.
The architecture of the LSM is relatively simple, consisting of an input layer, a large hidden layer (the reservoir), and a readout layer. 

\begin{figure}
    \centering
    \includegraphics[width=0.45\textwidth]{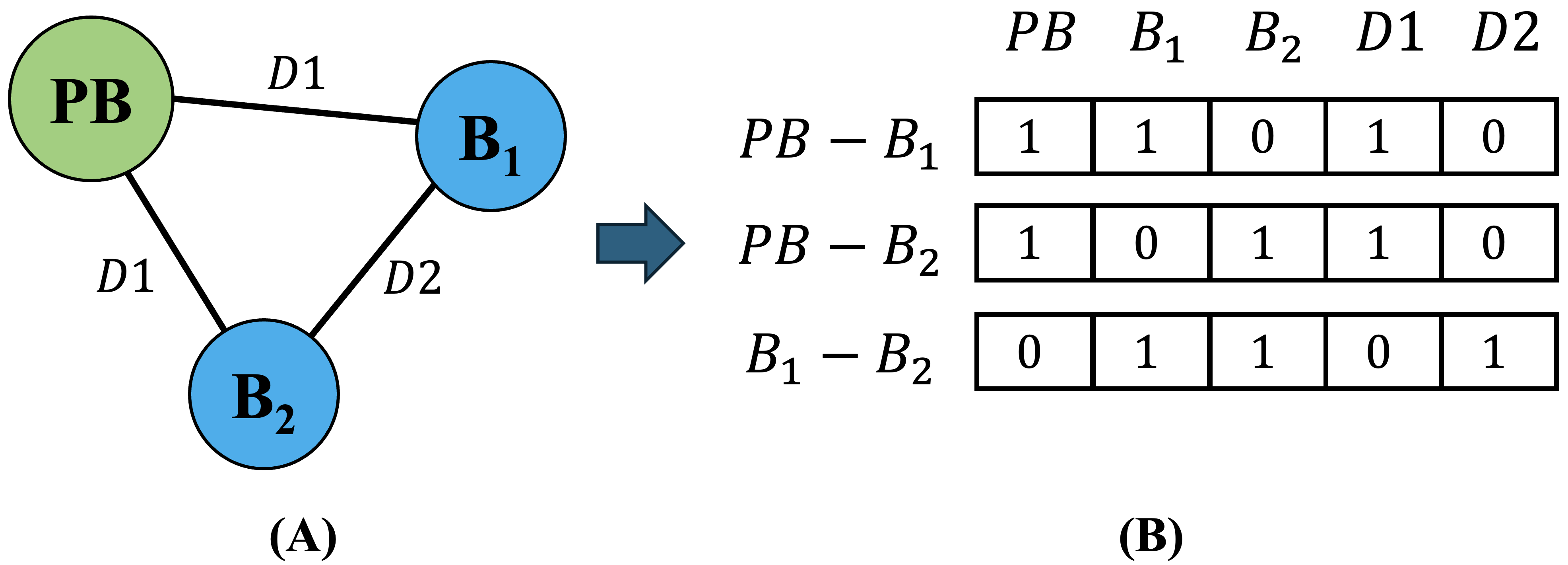}
    \caption{Encoding the structure of (A) PbB2 molecule into (B) three spike vectors.}
    \label{fig:pbb2}
\end{figure}

While spiking neurons have certain advantages, they come with the challenge of encoding graphical data into a set of spikes. In the case of atomic structures, a weighted undirected graph is used, with edge weights determined by the distance between the connected atoms. To encode this information, we generate a spike vector for each edge in the graph. Each node and distance value has a corresponding index in the vector. This index is set to 1 when the specific node or distance is involved in the edge, and 0 otherwise. 
%that would be 1 when represented and 0 otherwise.
As an illustrative example shown in Figure~\ref{fig:pbb2}, consider the structure of PbB2 where each atom is connected to the other to form a triangle. The lead atom (Pb) is connected to each boron (B) atom equidistantly at a distance of d1, and the boron (B) atoms are connected at a distance of d2. Therefore, the encoding would be three spike vectors: 

\begin{itemize}[leftmargin=4mm]
    \item A vector with a spike at node 1 (the lead atom), node 2 (one of the boron atoms), and d1 
    \item A vector with a spike at node 1, node 3 (the second boron atom), and d1
    \item A vector with a spike at node 2, node 3, and d2
\end{itemize}

After examining the dataset, we determined that a spike vector length of 165 would be sufficient. This vector encodes both node position (140 potential spikes) and distances between atoms (25 potential spikes, rounded to the nearest quarter angstrom from 0 to 6 angstroms). 
%After examining of the dataset, we determined that there would be 140 potential spikes to represent nodes and 25 to represent distances, for a total spike vector length of 165. Distances are rounded to the nearest quarter angstrom from 0 to 6. 
Due to the sparsity and size of this representation and contrary to standard GNN practice, we only utilize the initial unit cell of the atomic structure, rather than including periodic neighbor structures. This is also in line with \cite{b11} where non-periodic molecular structures were examined. 

\subsection{Hyperdimensional Computing}

\noindent
This section explores two approaches to study material science graph data:
1. Adapting GraphHD: We evaluate the applicability of the GraphHD methodology \cite{b12} for encoding material science graphs within the hyperdimensional computing (HDC) framework.
2. SSP-GrapHD: We further enhance the first approach by incorporating a recent strategy called Spatial Semantic Points (SSPs) to create 3D representations of the molecular structures.

\noindent
1. Adapting GraphHD:

\noindent
In GraphHD, the authors propose the following strategy:

%For the initial exploration of hyperdimensional computing, the methodology first presented in \cite{b12} was used. 
% In this work, the authors suggested the following strategy for encoding graph data: 
\begin{itemize}[leftmargin=4mm]
\item Symbolic Representaion: A random symbolic hypervector (denoted by H) is assigned to each potential node value, along with a separate vector (denoted by V) representing connected edges. Here, Multiply, Add, Permute (MAP) vectors with a value range of \{-1,1\} are chosen for both H and V. 
\item Node Memory Construction: A ``node memory" (NM) is constructed for each node, based on its connected neighbors and their corresponding edge weights:
\begin{equation}
    \mathit{NM}_i = \sum_{j} V_{w_{ij}} * H_j
\end{equation} 
Here, V is the edge hypervector that has been permuted based on the weight value W between the two nodes (effectively incorporating the weight information into the message passing process) and H is the connected neighbor node's hypervector. This operation is analogous to the message passing operation in GNNs. 
\item Node Embedding and Graph Representation: Each node representation (H) is then bound to its corresponding memory vector (NM) to create a final node embedding. All node embeddings are bundled together to form a representation of the entire graph:\begin{equation}
    G = \frac{1}{2}\sum_{i=1}^{n}H_i * \mathit{NM}_i
\end{equation}
\end{itemize}

To adapt this method to the materials science domain, we randomly initialize 118 hyperdimensional vectors, corresponding to the elements of the periodic table. We normalize the distance between atoms to values between 0 and 1 to align with the edge construction strategy presented in \cite{b12}. With these adjustments, the molecular structure graphs can be readily applied to the GrapHD methodology to construct hyperdimensional graph representations. Similar to the reservoir computing approach described earlier, we only consider the atomic unit cell structure. 

\vspace{5pt}
\noindent
2. SSP-GrapHD:

\noindent
To further explore the hyperdimensional computing space, we integrated another strategy called Spatial Semantic Pointers (SSP) \cite{b13}. SSPs offer a novel approach for representing continuous values, particularly spatial information. SSPs encode a point in space (s = [x, y]) using two hyperdimensional vectors, one for the x-axis (X) and one for the y-axis (Y). These vectors are unitary, meaning each vector has a magnitude of 1. 

The encoding process involves fractional binding of the axis vectors through a mathematical operation called circular convolution (denoted by $\otimes$). In simpler terms, this can be thought of as element-wise multiplication with a specific shift for each element. The resulting hypervector (S) represents the specific point in space:

\begin{equation}
        S = X^x \otimes Y^y.
    \end{equation}

An object, represented by a separate hyperdimensional vector (OBJ), can then be linked to its spatial location through binding. Furthermore, a ``spatial memory" (SM) can be constructed by bundling the object vectors with their corresponding spatial encodings (S):

\begin{equation}
        SM = \sum_{i=1}^{m} OBJ_i \otimes S_i
    \end{equation}

\noindent
Here, $OBJ_i$ represents the $i^{th}$ object and $S_i$ is its corresponding spatial location.

% first appeared in 2019 as a biologically plausible implementation of hyperdimensional vectors that could represent continuous values, in particular spatial information, through the following method: 
% \begin{itemize} [leftmargin=4mm]
%     \item Two unitary hyperdimensional vectors are initialized, one for the x axis (X) and one for the y axis (Y)
%     \item A point in space s = (x,y) can be encoded through fractional binding of the unitary axis vectors via circular convolution ($\otimes$): \begin{equation}
%         S = X^x \otimes Y^y.
%     \end{equation}
%     \item An object, represented by a hyperdimensional vector, can then be tied to a location through binding. From there, a spatial memory (SM) can be constructed through bundling:
%     \begin{equation}
%         SM = \sum_{i=1}^{m} OBJ_i \otimes S_i
%     \end{equation}
%     where OBJ is our object representation and S is its corresponding spatial location. 
% \end{itemize}

\begin{figure}
    \centering
    \includegraphics[width=0.48\textwidth]{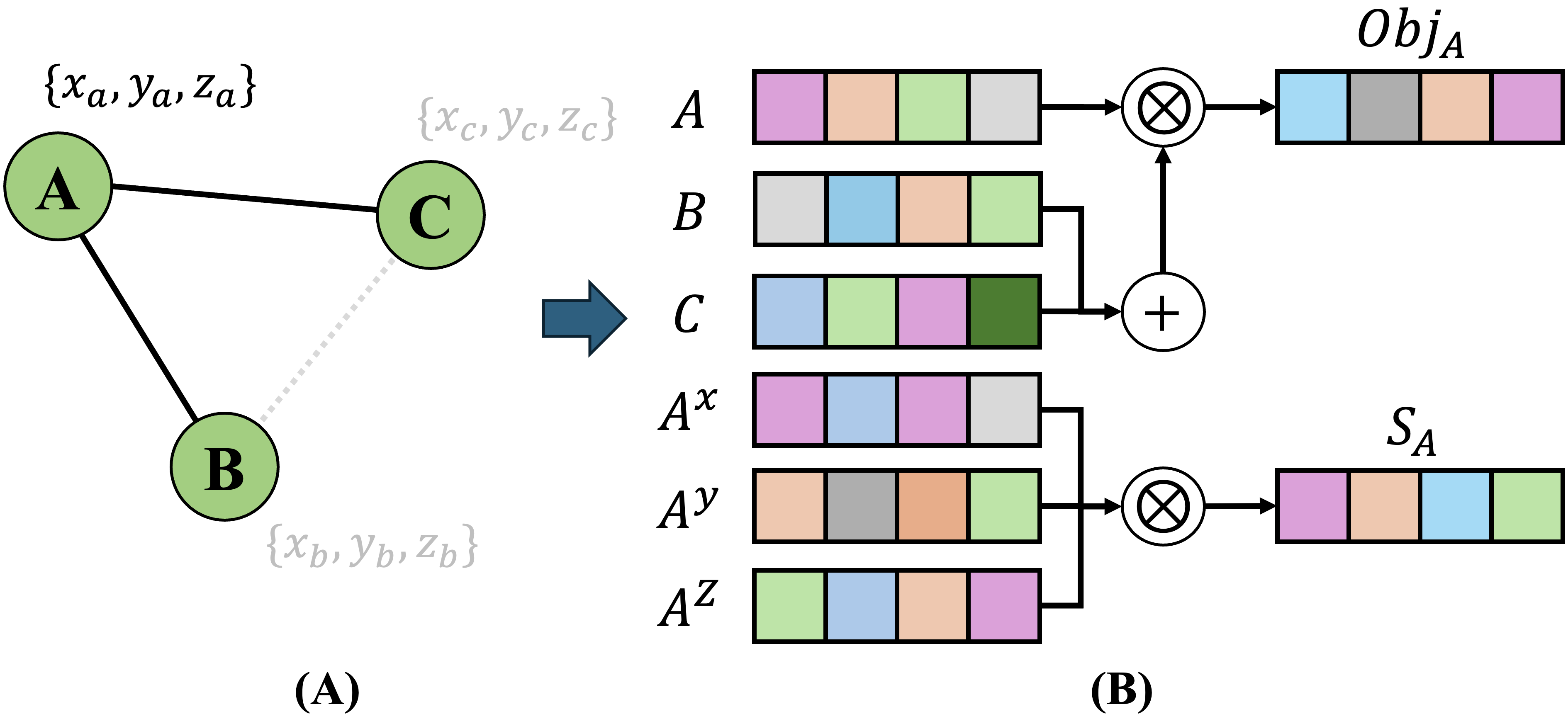}
    \caption{A simplified visualization of encoding (A) parent node A into (B) an object hyperdimensional vector that incorporates neighbor nodes and a spatial hyperdimensional vector that represents position. This process would be repeated for each node in the graph.}
    \label{fig:ssp-graphd-encoding}
\end{figure}

Therefore, we propose a novel approach called SSP-GrapHD that leverages the strengths of both methods to create a 3D spatial representation of our molecular structure, extending beyond a simple 2D graph. This approach combines the SSP equation for spatial encoding with a modified GraphHD equation:
% Therefore, there is potential to create a 3D spatial representation of our molecular structure that can extend beyond a simple 2D graph. To accomplish this, we can combine the SSP equation above with the GrapHD equation: 
\begin{equation}
    G = \frac{1}{2}\sum_{i=1}^{m}OBJ_i \otimes S_i
\end{equation}
Here, S is a spatial location, represented as above but with a third unitary vector to represent the Z axis. A diagram of this encoding process is highlighted within Figure~\ref{fig:ssp-graphd-encoding}. $OBJ_i$ now is a binding of an element representation with its corresponding memory vector $NM_i$ from the adapted GrapHD. However, SSP-GrapHD utilizes only an unweighted graph representation for node memory (NM)
%; only now, our memory M is represented via an unweighted graph: 
\begin{equation}
    \mathit{NM}_i = \sum_{j} H_j. 
\end{equation}  This is because our objects are tied to 3D locations in space, which inherently captures the distance information between them.  
% This combined approach is called \textit{SSP--GrapHD}.

%in the results section. 

\section{Results}

\noindent
For the reservoir approach, we initialized the weights across a random normal distribution, with a probability of connection between neurons based on their distance. The excitatory to inhibitory neuron ratio was 4:1. We explored reservoir sizes of 400, 1650, and 10,000. These reservoir sizes were selected based on existing literature on liquid state machines, while also considering the need to enforce sparsity in the network  \cite{b14}. The classification layer used a linear stochastic gradient descent classifier and a linear regressor was used for the regression task. Data was split 70\% for training and 30\% for testing. To account for the inherent randomness in reservoir computing, we averaged results over 25 independent runs. 

%with results averaged over 25 independent runs due to the random nature of the reservoir.

Our proposed approach, SSP-GrapHD, along with GrapHD, were evaluated using a hyperdimensional vector dimension size of 10,000 based on common practices in the literature. 
%Per popularity in the literature, both hyperdimensional computing strategies used a dimension size of 10,000. 
Both approaches utilized a stochastic gradient descent classifier and a linear regressor with a 70/30 train-test split. However, inspired by findings in \cite{b13}  suggesting neural networks' effectiveness with hyperdimensional representations, we also explored neural networks for the regression task. 
%indicate that neural networks could be highly effective at learning from hyperdimensional vector representations. Thus, neural networks were also examined for the regression task, 

For SSP-GrapHD, a single hidden layer network with a size of only 10 neurons achieved the best results.  In contrast, a two-hidden layer network performed best for GrapHD.

% with a two hidden layer network getting the best results for GrapHD and a single layer hidden network (with a size of only 10) getting the best results for the \textit{SSP--GrapHD}. 

We implemented both the reservoir and the SSP-GrapHD in Nengo \cite{b15} and GraphHD in torchHD \cite{b16}.

Table~\ref{tab1} summarizes the performance of all approaches on both the classification (accuracy) and regression (mean absolute error - MAE) tasks.
%All results can be seen in Table~\ref{tab1} for classification and regression. 
The GNN strategy ALIGNN \cite{b5} serves as the baseline.  ALIGNN is known for its effectiveness on various Materials Project tasks, including bandgap prediction.
%which achieves good performance across a number of The Materials Project tasks, including bandgap regression. 
It was trained and tested on the same data subset using hyperparameters recommended by the developers for small datasets. 
%Classification is reported with accuracy and regression with mean absolute error (MAE). 
For hyperdimensional strategies, neural network results for regression are reported with the linear regression results given in parenthesis. 

\begin{table}[htbp]
\caption{Results on The Materials Project data~\cite{b1} subset}
\begin{center}
\begin{tabular}{|c|c|c|}
\hline
\textbf{Method} & \textbf{MAE} & \textbf{Class. Acc.} \\
\hline
ALIGNN& 1.0688 &  -- \\
\hline
Reservoir--400 & 2.311 & 0.5330 \\
\hline
Reservoir--1650 & 1.7096 & 0.5084 \\
\hline
Reservoir--10k & 1.2035 & 0.5319 \\
\hline
GrapHD & 0.7025 (1.2270) & 0.7647 \\
\hline
SSP--GrapHD & 0.5181 (1.095) & 0.8235 \\
\hline
\end{tabular}
\label{tab1}
\end{center}
\end{table}

As can be seen in Table~\ref{tab1}, SSP-GrapHD achieves the best results by far for the data subset explored. Notably, SSP-GrapHD achieves a classification accuracy of $82.35\%$ and a mean absolute error (MAE) of 0.5181 on the regression task, outperforming all other approaches. 
Even without the neural network, a simple linear regressor approaches ALIGNN’s performance. These results indicate that the hyperdimensional computing strategies are capturing graph information very effectively, without any special deep learning architectures or feature engineering. The SSP-GrapHD results may also indicate that this strategy is very capable of generalizing, marking a potential area of interest for future efforts. 

The reservoir strategy, on the other hand, does not fair so well. Interestingly, while increasing the size of the reservoir does not help with classification, it appears to improve the regression task. Continuing to increase the reservoir size could lead to better representations; however, computational overhead begins to become an issue. Additionally, the reservoir's black box nature makes it difficult to tune, understand, and implement. 

\section{Conclusion and Future Work}
Hyperdimensional vectors have demonstrated several desirable qualities in this study: (1) they are largely transparent, (2) they are constructed through a series of simple algebraic operations, and (3) achieve good results. Further,  their potential implementation on neuromorphic hardware makes them computationally attractive. The results presented here, particularly the success of our proposed SSP-GrapHD approach, strongly indicate that hyperdimensional computing merits further exploration for material property prediction tasks.

The immediate next step is to evaluate the performance of hyperdimensional strategies on the full Materials Project dataset. This will provide a more comprehensive understanding of their effectiveness. Beyond the dataset size, several potential improvements can be explored. The current approach for node initialization could benefit from incorporating atomic properties more explicitly. Similar considerations apply to bond information.  Further, GraphHD currently only considers neighbors directly connected to a node for node memory. Like GNNs, it could potentially benefit from building node memory from neighbors further away. There are also other tasks beyond bandgap regression that could be examined in more detail. 

Interest in hyperdimensional vectors extends beyond the materials application as well. Future efforts will seek to compare hyperdimensional vectors more directly to traditional embedding strategies to better understand how well they incorporate information at a fundamental level. Exploration of using hyperdimensional vectors for representing other data types is also in progress, specifically for physics and image-based understanding. 

\section{Acknowledgment}
\noindent
The research was funded in part by National Science Foundation
through award CCF2319619.

\bibliographystyle{IEEEtran}
\bibliography{sample-base}

\begin{thebibliography}{10}
\providecommand{\url}[1]{#1}
\csname url@samestyle\endcsname
\providecommand{\newblock}{\relax}
\providecommand{\bibinfo}[2]{#2}
\providecommand{\BIBentrySTDinterwordspacing}{\spaceskip=0pt\relax}
\providecommand{\BIBentryALTinterwordstretchfactor}{4}
\providecommand{\BIBentryALTinterwordspacing}{\spaceskip=\fontdimen2\font plus
\BIBentryALTinterwordstretchfactor\fontdimen3\font minus \fontdimen4\font\relax}
\providecommand{\BIBforeignlanguage}[2]{{%
\expandafter\ifx\csname l@#1\endcsname\relax
\typeout{** WARNING: IEEEtran.bst: No hyphenation pattern has been}%
\typeout{** loaded for the language `#1'. Using the pattern for}%
\typeout{** the default language instead.}%
\else
\language=\csname l@#1\endcsname
\fi
#2}}
\providecommand{\BIBdecl}{\relax}
\BIBdecl

\bibitem{b1}
\BIBentryALTinterwordspacing
A.~Jain, S.~P. Ong, G.~Hautier, W.~Chen, W.~D. Richards, S.~Dacek, S.~Cholia, D.~Gunter, D.~Skinner, G.~Ceder, and K.~A. Persson, ``{Commentary: The Materials Project: A materials genome approach to accelerating materials innovation},'' \emph{APL Materials}, vol.~1, no.~1, p. 011002, 07 2013. [Online]. Available: \url{https://doi.org/10.1063/1.4812323}
\BIBentrySTDinterwordspacing

\bibitem{b2}
A.~Dunn, Q.~Wang, A.~Ganose, D.~Dopp, and A.~Jain, ``Benchmarking materials property prediction methods: the matbench test set and automatminer reference algorithm,'' \emph{npj Computational Materials}, vol.~6, no.~1, p. 138, Sep. 2020.

\bibitem{b3}
P.-P. De~Breuck, G.~Hautier, and G.-M. Rignanese, ``Materials property prediction for limited datasets enabled by feature selection and joint learning with {MODNet},'' \emph{npj Computational Materials}, vol.~7, no.~1, p.~83, Jun. 2021.

\bibitem{cong2023hyperparameter}
G.~Cong, S.~Kulkarni, S.~Lim, P.~Date, S.~Snyder, M.~Parsa, D.~Kennedy, and C.~Schuman, ``Hyperparameter optimization and feature inclusion in graph neural networks for spiking implementation,'' in \emph{2023 International Conference on Machine Learning and Applications (ICMLA)}, 2023, pp. 1541--1546.

\bibitem{b11}
S.~Wang, Y.~Li, D.~Wang, W.~Zhang, X.~Chen, D.~Dong, S.~Wang, X.~Zhang, P.~Lin, C.~Gallicchio, X.~Xu, Q.~Liu, K.-T. Cheng, Z.~Wang, D.~Shang, and M.~Liu, ``Echo state graph neural networks with analogue random resistive memory arrays,'' \emph{Nature Machine Intelligence}, vol.~5, no.~2, pp. 104--113, Feb. 2023.

\bibitem{b14}
L.~Deckers, I.~J. Tsang, W.~Van~Leekwijck, and S.~Latré, ``Extended liquid state machines for speech recognition,'' \emph{Frontiers in Neuroscience}, vol.~16, 2022.

\bibitem{b5}
K.~Choudhary and B.~DeCost, ``Atomistic line graph neural network for improved materials property predictions,'' \emph{npj Computational Materials}, vol.~7, no.~1, p. 185, Nov. 2021.

\bibitem{b4}
T.~K. Rusch, M.~M. Bronstein, and S.~Mishra, ``A survey on oversmoothing in graph neural networks,'' 2023.

\bibitem{b6}
\BIBentryALTinterwordspacing
R.~Ruff, P.~Reiser, J.~Stühmer, and P.~Friederich, ``Connectivity optimized nested line graph networks for crystal structures,'' \emph{Digital Discovery}, vol.~3, pp. 594--601, 2024. [Online]. Available: \url{http://dx.doi.org/10.1039/D4DD00018H}
\BIBentrySTDinterwordspacing

\bibitem{b8}
\BIBentryALTinterwordspacing
G.~Cong, S.-H. Lim, S.~Kulkarni, P.~Date, T.~Potok, S.~Snyder, M.~Parsa, and C.~Schuman, ``Semi-supervised graph structure learning on neuromorphic computers,'' in \emph{Proceedings of the International Conference on Neuromorphic Systems 2022}, ser. ICONS '22.\hskip 1em plus 0.5em minus 0.4em\relax New York, NY, USA: Association for Computing Machinery, 2022. [Online]. Available: \url{https://doi.org/10.1145/3546790.3546821}
\BIBentrySTDinterwordspacing

\bibitem{b9}
D.~Dold and J.~S. Garrido, ``Spike: spike-based embeddings for multi-relational graph data,'' in \emph{2021 International Joint Conference on Neural Networks (IJCNN)}, 2021, pp. 1--8.

\bibitem{b10}
\BIBentryALTinterwordspacing
V.~C. Chian, M.~Hildebrandt, T.~Runkler, and D.~Dold, ``Learning through structure: Towards deep neuromorphic knowledge graph embeddings,'' in \emph{2021 International Conference on Neuromorphic Computing (ICNC)}.\hskip 1em plus 0.5em minus 0.4em\relax IEEE, Oct. 2021. [Online]. Available: \url{http://dx.doi.org/10.1109/ICNC52316.2021.9607968}
\BIBentrySTDinterwordspacing

\bibitem{b12}
\BIBentryALTinterwordspacing
P.~Poduval, H.~Alimohamadi, A.~Zakeri, F.~Imani, M.~H. Najafi, T.~Givargis, and M.~Imani, ``Graphd: Graph-based hyperdimensional memorization for brain-like cognitive learning,'' \emph{Frontiers in Neuroscience}, vol.~16. [Online]. Available: \url{https://par.nsf.gov/biblio/10338293}
\BIBentrySTDinterwordspacing

\bibitem{b13}
B.~Komer, T.~C. Stewart, A.~Voelker, and C.~Eliasmith, ``A neural representation of continuous space using fractional binding.'' in \emph{CogSci}, 2019, pp. 2038--2043.

\bibitem{b15}
T.~Bekolay, J.~Bergstra, E.~Hunsberger, T.~DeWolf, T.~C. Stewart, D.~Rasmussen, X.~Choo, A.~R. Voelker, and C.~Eliasmith, ``Nengo: a python tool for building large-scale functional brain models,'' \emph{Frontiers in neuroinformatics}, vol.~7, p.~48, 2014.

\bibitem{b16}
M.~Heddes, I.~Nunes, P.~Verg{\'e}s, D.~Desai, T.~Givargis, and A.~Nicolau, ``Torchhd: An open-source python library to support hyperdimensional computing research,'' \emph{arXiv preprint arXiv:2205.09208}, 2022.

\end{thebibliography}

\end{document}